A New $k$-Shortest Path Search Approach based on Graph Reduction

Yasuo Yamane and Hironobu Kitajima

Fujitsu Limited, 17-25, Shinkamata 1-chome, Ota-ku, Tokyo, 144-8588, Japan

Abstract

We present a new approach called GR (Graph Reduction) algorithm for searching loop-less $k$-shortest paths (1st to $k$-th shortest paths) in a graph based on graph reduction. Let a source vertex and a target vertex of $k$-shortest paths be $v_s$ and $v_t$ respectively. First our approach computes shortest paths to every vertex from $v_s$ and $v_t$ respectively, and reduce a graph to a subgraph that contains all vertices and edges of loop-less $k$-shortest paths using the already computed shortest paths, and apply an existing $k$-shortest path search algorithm to the reduced graph. A graph can be reduced quickly after computing the shortest paths using them, therefore a very efficient search can be achieved.

In an experiment using a hypercube graph which has $16384$ vertices where $k = 128$, the number of vertices is reduced to about 1/22, and a variant of Dijkstra's algorithm for k-shortest path search were speeded up by about 365 times. We implemented a fast k-shortest path variant of bidirectional Dijkstra's algorithm ($k$-biDij) which is the state-of-the-art algorithm and the fastest as long as we know, GR outperforms $k$-biDij in dense scale-free graphs. However, $k$-biDij outperforms GR in hypercube-shaped and sparse scale-free graphs, but even then GR can also speed up it by 12.3 and 2.0 times respectively by precomputing all-pairs shortest paths.

We also show the graph reduction can be done in time complexity $O(m + n \log n)$.

We also introduce our improvements to $k$-biDij simply.

1. Introduction

In this paper, we present a new approach called GR (Graph Reduction) algorithm for searching loop-less $k$-shortest paths (1st to $k$-th shortest paths) in a graph based on graph reduction. GR algorithm is called GR simply below.

Searching $k$-shortest paths has been becoming important for analyzing graphs. For example, getting the shortest path between two vertices is important for knowing the relationship between them, but we can know more detailed relationships by knowing $k$-shortest paths between them. However, $k$-shortest path search costs more than shortest path search, and the size of the analyzed graphs has becoming larger. Therefore, speeding up the $k$-shortest path search is important to get an timely analysis results.

In $k$-shortest path search, there are two major ways: One permits including loop and the other not.

The latter loop-less way needs more time including ours.

We introduce conventional loop-less algorithms simply: A famous one is Yen's algorithm [Yen71]. A $k$-shortest path variant of Dijkstra method [$k$-shortest] is known (called $k$-Dijkstra algorithm, or $k$-Dij in this paper). A bidirectional variant of Dijkstra algorithm is known, and a $k$-shortest path variant of this was proposed without detailed description [Zhao14] (called $k$-bidirectional Dijkstra algorithm, or shortly $k$-biDij in this paper). We implemented this and improved in some ways and used it as one of conventional algorithms in our experiments for comparison, which is the fastest as long as we know. We introduce our improvements to $k$-biDij simply in the appendix.

Next, we introduce GR briefly. Let a source (starting) and a target (terminating) vertex of $k$-shortest paths to be computed be $v_s$ and $v_t$ respectively. Fist GR computes the shortest paths to every vertex from $v_s$ and $v_t$ respectively (These shortest paths are called "st-shortest paths" below where st means source and target). Secondly the graph is reduced to a subgraph that contains all the vertices and the edges of the $k$-shortest paths to be computed using the st-shortest paths. Lastly one of existing $k$-shortest path search algorithms is applied to the subgraph. By this approach we can achieve a very efficient search because a graph can be reduced quickly after computing the st-shortest paths using them. GR is a meta algorithm, or an operator in the sense that any existing algorithm can be applied, that is, become a parameter of GR. GR algorithm using an existing algorithm $E_x$ is expressed by GR($E_x$) below.

In our experiments, we compared $k$-Dij with GR($k$-Dij) and $k$-biDij with GR($k$-biDij) respectively by measuring their CPU time. Hypercube-shaped graphs, sparse and dense scale-free graphs were used.

In an experiment using a hypercube-shaped graph which has 16384 vertices where $k = 128$, the number of vertices is reduced to about 1/22, and $k$-Dij was speeded up by about 365 times. GR outperforms k-biDij in dense scale-free graphs. On the other hand, $k$-biDij outperforms GR in hypercube-shaped and sparse scale-free graphs. However, even then GR can also speed up it by about 12.3 and 2.0 times respectively by precomputing all-pairs shortest paths.

We also show the graph reduction can be done in time complexity $O(m + n \log n)$, which is not dependent on $k$.

2. Relative Works

We explain conventional algorithms in more detail in this section.

1) Yen's algorithm [Yen71]

As mentioned above, this is a famous algorithm to compute loop-less $k$-shortest paths. However we excluded this in our experiments because $k$-Dij seemed to outperform it in our experiments.

2) $k$-Dijkstra algorithm [K-shortest]

Although Dijkstra's algorithm is famous as a fast one for computing all the shortest paths from a source vertex to all the other vertices, this algorithm can also be used to compute the shortest path to a target vertex by stopping as soon as it reaches the vertex.

A variant of the latter Dijkstra's algorithm generalized for $k$-shortest paths search is introduced in [k-shortest], which we call "$k$-Dijkstra algorithm". The version in [k-shortest] seems to be an algorithm for searching $k$-shortest paths with loops. It searches beyond a vertex $v$ if the number of paths from the source vertex is less than or equal to $k$. We modified it so that it can check whether the new path which has reached $v$ contains $v$ on the way, that is, a loop, and it can search beyond $v$ if the new path does not contain such a loop. It worked correctly for searching loop-less $k$-shortest paths. We also added some improvements mentioned in Appendix 1) and 2). We used this version for our experiments and call it as $k$-Dijkstra algorithm or $k$-Dij simply below.

3) $k$-bidirectional Dijkstra algorithm [Zhao14]

Although Dijkstra's algorithm searches starting from a source vertex, bidirectional Dijkstra's algorithm searches starting from both a source vertex and target vertex and it is said to be faster in many cases. It intuitively seems that searched areas are reduced by starting both vertices and searching synchronizing.

Extending bidirectional Dijkstra algorithm to $k$-shortest path search was proposed without detailed description as mentioned above. We call this algorithm $k$-bidirectional Dijkstra algorithm or $k$-biDij simply below. We implemented this algorithm and speeded it up by adding improvements as follows:
   a) reducing the number of paths enqueued into priority queues
   b) After $k$ paths (not shortest) from the source vertex and the target vertex are got, utilizing the length of the longest path of them for pruning.

The details are mentioned in the appendix. Below $k$-biDij means this implementation.

3. GR Algorithm

In this section, we explain GR algorithm in more detail. 3.1 shows our basic idea, and 3.2 shows the details of the algorithm.

We explain our algorithm using undirected weighted graphs to simplify our explanations, but it is easy to modify it for them to be used in directed graphs or unweighted graphs as mentioned later.

3.1 Our basic idea

The core of our idea is to reduce the original graph to a subgraph which contains the $k$-shortest paths to be solved. If the graph can be reduced quickly and the reduced subgraph is small enough, we can get the solution, the $k$-shortest paths, very fast by applying the existing fast algorithm to the reduced one. The more detailed summary of our idea are as follows:

Let all the vertices be $v_1, v_2, \cdots, v_n$, the source vertex of $k$-shortest paths be $v_s$ and the target vertex of them be $v_t$. The st-shortest paths, that is, the shortest paths between $v_s$ or $v_t$ and every vertex are computed, and the distances of them are stored as vectors. For example, the distances from $v_s$ to every vertices is expressed by a vector of size $n$ $[d_i]$ where $d_i$ is the distance from $v_s$ to $v_i$.

It is well-known that the shortest paths from a vertex, $v$ to every vertex can be expressed by a tree whose root is $v_i$, which is called "shortest path tree". The shortest paths are stored as vectors. For example, the shortest paths from $v_s$ to every vertex is expressed by a vector of size $n$ $[p_i]$ where $p_i$ is the parent of $v_i$ on the shortest path from $v_s$ to $v_i$.

We explain the meanings of the symbols used below as follows: We express a graph to be searched by $G = (V, E)$ where $V$ is a set of vertices and $E$ is a set of edges. Let $n = |V|$. The shortest path from $v_s$ to $v_i$, and that from $v_i$ to $v_t$ which are mentioned above are expressed by $sp(v_s, v_i)$ and $sp(v_i, v_t)$ respectively. The length (distance) of a path $p$ is expressed by $d(p)$. The distance between $v_i$ and $v_j$, that is, the length of the shortest path between $v_i$ and $v_j$, are expressed by $d(v_i, v_j)$, that is, $d(v_i, v_j) = d_{ij}$. The length of $i$-th shortest path from $v_s$ to $v_t$ is expressed by $spd_i$. Then the condition that a path $p$ is a solution can be said to be that $p$ is loop-less and $d(p) \leq spd_k$.

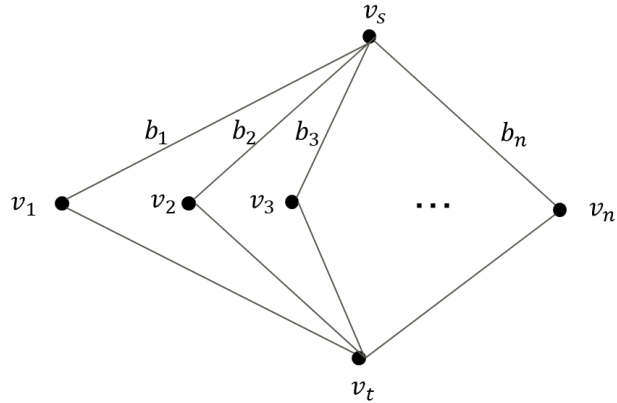

Fig.3.1 by-way-of shortest paths

As shown in Fig.3.1, let us consider the path from $v_s$ to $v_t$ by way of each $v_i$ where both the path from $v_s$ to $v_i$ and the path from $v_i$ to $v_t$ are the shortest paths. We call this path a "by-way-of shortest path", and express it by $b_i$ or $b(v_i)$. It holds true that $b(v_i)$ is the shortest among all the paths from $v_s$ to $v_t$ by way of $v_i$, which is used later, because if there is a shorter one, it is contrary to the fact that the path from $v_s$ to $v_i$ and that from $v_i$ to $v_t$ on $b_i$ are the shortest.

Here, $b_s$ and $b_t$ are the shortest path from $v_s$ to $v_t$.

The essence of our idea is as follows: The by-way-of shortest paths $b_1, b_2, \cdots, b_n$ are easily computed by connecting $sp(v_s, v_i)$ and $sp(v_i, v_t)$ after st-shortest paths are computed. They can be redundant, that is, for some $b_i$ and $b_j$ $(i \neq j)$ the path of $b_i$ and that of $b_j$ can coincide completely. At first the redundant ones are transformed into unique ones. The remaining unique by-way-of shortest paths are sorted in increasing order of $d(b_i)$. Let the result be $b_{i_1}, b_{i_2}, \cdots, b_{i_{n'}} (n' \leq n)$. Scanning from $b_{i_1}$, $k$ loop-less by-way-of shortest paths are tried to find. Let us consider they are found. In case the number of them is less than $k$, we call it "*insufficient case*" later. In this case, we cannot reduce the graph, and let $V' = V$.

Let $b_{i_e}$ be the $k$-th loop-less by-way-of shortest path. Then, letting the original graph be $G = (V, E)$, $V'$ be the set of vertices contained in $b_{i_1}, b_{i_2}, \cdots, b_{i_e}$, and $E' = \{(v,w)|(v,w) \in E, and\ v, w \in V'\}$, a subgraph $G' = (V', E')$ is the reduced one containing the $k$-shortest paths to be solved. The correctness is proved in the theorem 1.

Here, you should pay attention to the following points: Let us consider cases where $e < n'$, that is, a graph can be reduced. There are $k$ loop-less unique by-way-of shortest paths from $v_s$ to $v_t$ in $b_{i_1}, b_{i_2}, \cdots, b_{i_e}$. The first point is that it is possible they are not the $k$-shortest paths to be solved, that is, $spd_k \leq d(b_{i_e})$. This also guarantees that $k$-shortest paths can be computed.

The second point is that for some $j \leq e$ $b_{i_j}$ can contain a loop. We thought $b_{i_j}$ should be removed at first, but it failed. We explain why graph reduction does not work correctly then in the following Lemma.

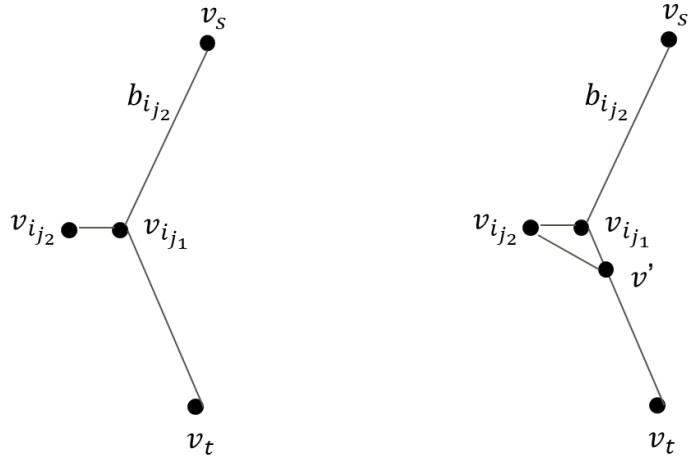

(a) by-way-of shortest path with a loop    (b) very close loop-less shortest path

Fig.3.2 by-way-of shortest path with a loop which should not be removed

Lemma

If $i_j < i_e$ and $b_{i_j}$ contains a loop, the vertices in a loop must be added to $V'$.

Proof:

Let us consider a by-way-of shortest path $b_{i_{j_2}} (j_2 < e)$ containing a loop letting $v_{i_{j_1}}$ and $v_{i_{j_2}}$ be adjacent as shown in Fig.3.2 (a) in case that a graph can be reduced. We can let $j_2 < e$ because $b_{i_{j_e}}$ does not contain a loop. The vertex $v_{i_{j_1}}$ is shared by both $sp(v_s, v_{i_{j_2}})$ and $sp(v_{i_{j_2}}, v_t)$, so $v_{i_{j_1}}$ and $v_{i_{j_2}}$ form a loop. That is, $b_{i_{j_2}}$ is equal to $b_{i_{j_1}}$ to which $v_{i_{j_2}}$ was added. Therefore $d(b_{i_{j_1}}) < d(b_{i_{j_2}})$, and $i_{j_1} < i_{j_2}$. Let $d(v_{i_{j_1}}, v_{i_{j_2}})$ be extremely small. In addition, as shown in Fig.3.2(b), let us assume that there is a vertex $v'$ is on $sp(v_{i_{j_2}}, v_t)$, and $d(v_{i_{j_2}}, v')$ is extremely small. Then letting $p$ be the path from $v_s$ to $v_t$ through $v_{i_{j_1}}, v_{i_{j_2}}$ and $v'$, $d(p)$ is slightly larger than $d(b_{i_{j_1}})$. Here letting us assume $d(b_{i_{j_2+1}})$ is sufficiently larger than $d(b_{i_{j_1}})$, $d(b_{i_{j_2}}) < d(b_{i_{j_2+1}})$. This means $d(p) < d(b_{i_{j_e}})$, that is, $d(p)$ is smaller than $k$-th by-way-of shortest path. So $v_{i_{j_2}}$ should be put into $V'$, because the all the vertices of $b_{i_{j_e}}$ are put into $V'$. The above is why the by-way-of shortest paths with loops should not be removed.    Q.E.D.

Theorem 1

   Graph $G' = (V', E')$ contains the $k$-shortest paths to be solved.

Proof:

   In the "insufficient case", we let $V' = V$, so we have only to prove in the other cases. Let the set of vertices contained in $b_{i_1}, b_{i_2}, \cdots, b_{i_e}$ be $V'$ and let us sort the loop-less unique by-way-of shortest paths in $b_{i_1}, b_{i_2}, \cdots, b_{i_e}$ in increasing order of $d(b_{i_1})$ and let the result be $b_{j_1}, b_{j_2}, \cdots, b_{j_k}$. As mentioned before, $spd_k \leq d(b_{j_k})$.

   Here, let us consider vertex $v$ which is not contained in $V'$. If it is said that there is not such a path $p$ from $v_s$ to $v_t$ by way of $v$ as $d(p) < spd_k$, it can be said that it is not necessary to add $v$ to $V'$, so we can finish the proof.

   Let us consider a by-way-of shortest path $b$ by way of $v$. As mentioned before, $d(b) \leq d(p)$. And it holds true that $d(b_{j_k}) \leq d(b)$ because if $d(b) < d(b_{j_k})$, $b$ should be one of $b_{j_1}, b_{j_2}, \cdots, b_{j_k}$. So $spd_k \leq d(b)$ because $spd_k \leq d(b_{j_k})$ as mentioned above, and $spd_k \leq d(p)$. So there is not such a path $p$ from $v_s$ to $v_t$ by way of $v$ as $d(p) < spd_k$.    Q.E.D.

3.2 The details of GR algorithm

   We show two versions of GR algorithm. One is a primitive one which realizes the idea mentioned above primitively. This is primitive and understandable, but it seems a little bit heavy to eliminate redundancy of by-way-of shortest paths. The other is a speeded-up one which can reduce the graph faster because it is not necessary to remove redundancy of by-way-of shortest paths.

3.2.1   Primitive GR algorithm

   The primitive GR algorithm is shown below

---

primitive GR algorithm

   input: $G = (V, E)$, $E_x$  $k$, $v_s$, $v_t$

output: $k$-shortest paths

1) Compute the shortest paths from $v_s$ and $v_t$ to every vertex.
2) Compute the set of by-way-of shortest paths $B = \{b_1, b_2, \cdots, b_n\}$ using the result of 1).
3) Remove redundancy of $B$ (for example, using sorting or hashing), and let the resulting set be $B_u = \{b_{h_1}, b_{h_2}, \cdots, b_{h_{n'}}\}$ $(n' \leq n)$, where $u$ of $B_u$ means unique.
4) Sort $B_u$ in increasing order of $d(b_{h_i})$, and let the resulting sequence be $B_s = \{b_{i_1}, b_{i_2}, \cdots, b_{i_{n'}}\}$ $(n' \leq n)$.
5) Find the loop-less k-th by-way-of shortest path $b_{i_e}$ scanning $B_s$ from $b_{i_1}$. If not found, let $G' = G$ and go to 8.
6) Let the set of vertices contained in $b_{i_1}, b_{i_2}, \cdots, b_{i_e}$ be $V'$.
7) Let $E' = \{(v, w) | (v, w) \in E, and\ v, w \in V'\}$. Then, a Graph $G' = (V', E')$ is the subgraph which contains $k$-th shortest paths to be solved.
8) Compute the k-shortest paths by applying an existing algorithm $E_x$ to $G'$.

### 3.2.2 Speeded-up GR algorithm

Speeded-up GR algorithm is shown below, which differs from the primitive one in 3) to 15). A variable $c$ is a counter to count the number of loop-less by-way-of shortest paths.

speeded-up GR algorithm
    input: $G = (V, E)$, $E_x$, $k$, $v_s$, $v_t$
    output: $k$-shortest paths

1) Compute the shortest paths from $v_s$ and $v_t$ to every vertex.
2) Compute the set of by-way-of shortest paths $B = \{b_1, b_2, \cdots, b_n\}$ using the result of 1).
3) Sort $B$ in increasing order of $d(b_i)$, and let the resulting sequence be $B_s = \{b_{i_1}, b_{i_2}, \cdots, b_{i_n}\}$.
4) $V' = \{\ \}; c = 0$
5) $while\ B_s \neq \{\ \}$:
6)     Let the first of $B_s$ be $b_{i_j}$.
7)     $if\ b_{i_j}$ does not contain a loop:
8)         Let the vertices contained in $b_{i_j}$ be $v_{j_1}, v_{j_2}, \cdots, v_{j_h}$.
9)         $V' = V' \cup \{v_{j_1}, v_{j_2}, \cdots, v_{j_h}\}$
10)        $B_s = B_s - \{b_{j_1}, b_{j_2}, \cdots, b_{j_h}\}$
11)        $c = c + 1$
12)        $if\ c = k: break$

13)    *else*:

14)         $V' = V' \cup \{v_{i_j}\}$

15)         $B_s = B_s - \{b_{i_j}\}$

16) Let $E' = \{(v,w)|(v,w) \in E, and\ v, w \in V'\}$. Then, a Graph $G' = (V', E')$ is the subgraph which contains $k$-th shortest paths to be solved.

17) Compute the k-shortest paths by applying an existing algorithm $E_x$ to $G'$.

---

We prove why this algorithm can get the same result as that of the primitive GR algorithm. The part from 8) to 12) corresponds to the part of removing redundancy of by-way-of shortest paths in the primitive one, and the part of 14) and 15) corresponds to adding the vertices in the by-way-of shortest paths having loops to $V'$. It seems that you can understand the reason if you understand these two parts.

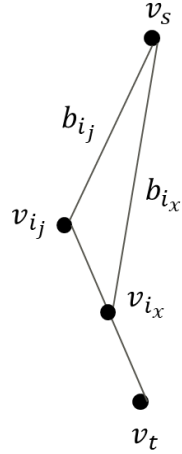

Fig. 3.3 Vertex $v_{i_x}$ and by-way-of shortest paths $b_{i_j}$、$b_{i_x}$

Theorem 2

The speed-up algorithm works validly.

Proof:

First, we explain the former. Let the vertices contained in $b_{i_j}$ be $v_{j_1}$, $v_{j_2}$, $\cdots$, $v_{j_h}$. As $b_{i_j}$ is the by-way-of shortest path by way of $v_{i_j}$, $b_{i_j}$ is equal to one of $v_{j_1}$, $v_{j_2}$, $\cdots$, $v_{j_h}$. Let $v_{j_1}$ be the same as $v_{i_j}$, so $j_1 = i_j$. Although $v_s$ and $v_t$ are one of $v_{j_1}$, $v_{j_2}$, $\cdots$, $v_{j_h}$ respectively, let us consider the vertex $v_{i_x}$ shown in Fig.3.3 which is neither $v_{j_1}$ nor $v_s$ nor $v_t$. Then $d(b_{i_j}) = d(b_{i_x})$. The reason is as follows: As $b_{i_j}$ the first of $B_s$, $d(b_{i_j}) \leq d(b_{i_x})$. If $d(b_{i_j}) < d(b_{i_x})$, The distance from $v_s$ to $v_{i_x}$ on $b_{i_x}$ is longer than that from $v_s$ to $v_{i_x}$ on $b_{i_j}$. This is contrary to that the distance from $v_s$ to $v_{i_x}$ on $b_{i_x}$ is the shortest. So $d(b_{i_j}) = d(b_{i_x})$. This means it is possible or allowed that $b_{i_x}$ is the same as $b_{i_j}$ because we can make the path from $v_s$ to $v_{i_x}$ on $b_{i_x}$ be the same as the path from $v_s$ to $v_{i_x}$ through $v_{i_j}$ on $b_{i_j}$ when selecting shortest paths in the process of making st-shortest paths. That is, it means $b_{i_x}$ can be removed regarding it as the same one as $b_{i_j}$.

Secondly, we explain the latter. In 14), only $v_{i_j}$ is added to $V'$. The question we have to consider here is whether the vertices of $b_{i_x}$ and $v'$ should be added to $V'$, in such a case as shown in Fig.3.4 (a), that is, some $v'$ other than $v_{i_j}$ is contained in the loop. On $v'$, it is unnecessary because as $d(b(v')) < d(b_{i_j})$, $v'$ is already added to $V'$. Although the loop in Fig.3.4 (a) is linear, the same thing holds true in such a case as a cyclic loop. Because as the length of the path from $v_{i_x}$ to $v_{i_j}$ through $v'$ and the length of the path from $v_{i_x}$ to $v_{i_j}$ through $v''$ are the same, $d(b(v')) < d(b_{i_j})$ and $d(b(v'')) < d(b_{i_j})$. On the vertices of $b_{i_x}$, the vertices of $b_{i_x}$ is already added to $V'$ because $d(b_{i_x}) < d(b_{i_j})$.    Q.E.D.

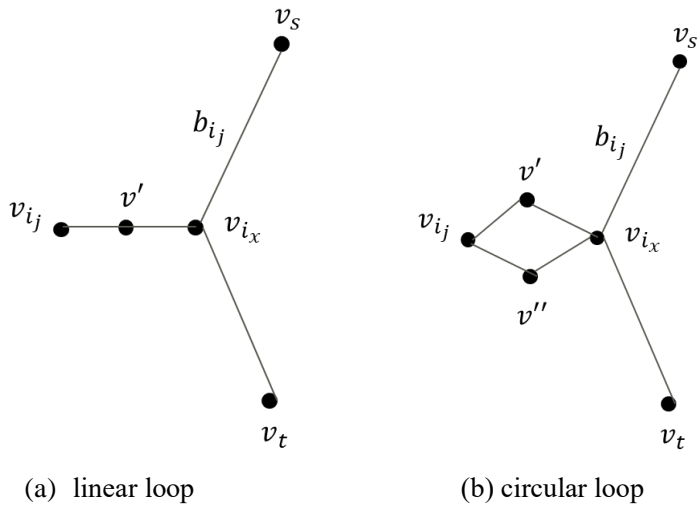

(a)  linear loop            (b) circular loop

Fig.3.4 The case where $b_{i_j}$ contains a loop

Below let GR algorithm mean the speeded-up one.

The speed-up GR algorithm is not only faster than the primitive one, the size of the reduced subgraph is less than or equal to that of the primitive one. The following theorem proves that.

Theorem 3

The size of the reduced subgraph of the speed-up GR algorithm is less than or equal to that of the primitive one.

Proof:

In by-way-of shortest paths which have loops, the number of vertices of the speed-up algorithm is the same as that of the primitive one. This is obvious from the proof of Theorem 2. The number of vertices can be different in the loop-less by-way-of shortest paths. The reason is as follows: In Fig.3.3, $b_{i_j}$ and $b_{i_x}$ are regarded as the same ones in the speed-up algorithm, but in the primitive one, they can be different as shown in Fig. 3.3 depending on how to generate st-shortest paths. Therefore, in the part of loop-less by-way-of shortest paths, the number of vertices of the speed-up GR algorithm is less than or equal to that of the primitive one.     Q.E.D.

3.3 Extension to other types of graph

Although GR algorithm mentioned above can be used for undirected weighted graphs, we explain the modifications for directed and unweighted graphs.

The modification for directed graphs is only as follows: When creating by-way-of shortest paths, in case that for some $v_i$ it is possible that there is either no path from $v_s$ to $v_i$ or no path from $v_i$ to $v_t$. Then it is necessary to odify GR algorithm so that $b_i$ may be removed from $B$.

In case of unweighted graphs, we have only to let all weights be 1.

4．Evaluation

We compared GR with $k$-Dij and $k$-biDij where we let $k = \sqrt{n}$. We excluded Yen because $k$-Dij outperforms it. The edge lengths were randomly chosen between 0 and 1. We used two kinds of graphs as follows:

1) hypercube-shaped graph

We measured in cases where $n = 2^{2i}$ ($i = 3,4,5,6,7$), that is, $n = 64, 256, 1024, 4096$, and $16384$. The degree of each vertex of the hypercube is $\log_2 n$, and they are $6, 8, 10, 12$ and $14$ respectively for $n = 64, 256, 1024, 4096$, and $16384$. We chose the two furthest vertices as $v_s$ and $v_t$ regarding the graph as an unweighted graph. To be exact, we should choose some different pairs of $v_s$ and $v_t$ randomly because the behavior of algorithms can differ depending on the paths.

2) scale -free graph

Likewise, we measured in cases where $n = 64, 256, 1024, 4096$, and $16384$. The graph is created as follows: When creating a graph having $n$ vertices, first a complete graph having $n'$ vertices where $n' < n$ is created. Secondly the remaining $n - n'$ vertices are added one by one as follows: Let one of them be $v$. Let randomly chosen $n'$ vertices be $v_1, v_2, \cdots, v_{n'}$. Then Let $V = V \cup \{v\}$ and $E = E \cup \{(v, v_1), (v, v_2), \cdots, (v, v_{n'})\}$. The probability of choosing $v_i$ ($i = 1, 2, \cdots, n'$) was let to be proportional to the degree of $v_i$. We chose the two vertices whose identifiers are $0$ and $n - 1$ as $v_s$ and $v_t$. To be exact, we should choose some different pairs of $v_s$ and $v_t$ randomly because the distance between $v_s$ and $v_t$ can be too short, although the results seem to be appropriate.

For measurement, we used FUJITSU Workstation CELSIUS M740 with Intel Xeon E5-1603 v4 (2.80GHz) and 32GB main memory, programing language Python, and OS Linux.

4.1 Comparison in hypercube-shaped graphs

Fig. 4.1 and Table 4.1 shows the comparison in CPU time of four algorithms. Fig. 4.1 is shown in a double-logarithmic graph. CPU time graphs are shown in the same manner below. The column whose name is $k$-biDij/GR($k$-biDij) shows the rate of the CPU time of $k$-biDij against that of GR($k$-biDij), that is, how many times GR($k$-biDij) is faster than $k$-biDij. When it is smaller than 1.0, that means that GR($k$-biDij) is outperformed. The column of $k$-Dij/GR($k$-Dij) is likewise. We call these values "speed rate".

When $n = 16384$, GR($k$-biDij) is overperformed by $k$-biDij by about 2 times, and it is difficult to judge whether the speed rate increases or decreases as $n$ increases from only these data. GR($k$-Dij) outperforms $k$-Dij by 365 times, and the speed rate increases as $n$ is increases.

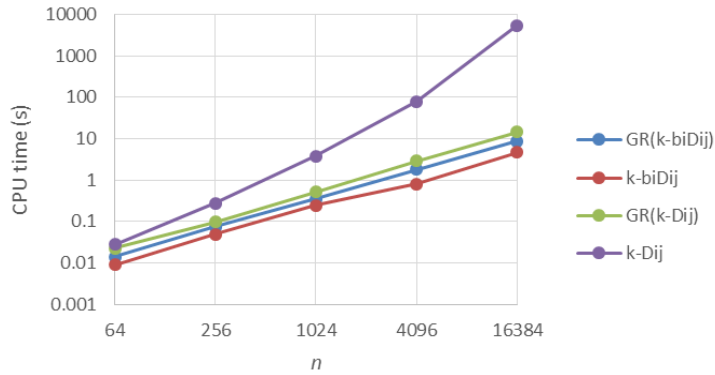

Fig. 4.1 Comparison in CPU time four algorithms in hypercube-shaped graphs

Tables 4.2 and 4.3 show the breakdown of CPU time of GR($k$-biDij) and GR($k$-Dij) and the number of vertices of the reduced subgraph and its rate respectively. GR algorithm consists of two parts: One is a part where a reduced subgraph which contains $k$-shortest paths is generated, and the other is a part where an existing $k$-shortest path search is applied to the subgraph. The former part consists of two parts, one where st-shortest paths are computed and the other where a graph is reduced utilizing computed st-shortest paths. The second part is called "proper graph reduction" or "proper reduction" in the sense that a graph is reduced in a proper meaning. The breakdown of CPU time shows the CPU time of the three parts, which corresponds to "st-shortest paths", "proper reduction", and "subgraph search". Each column whose name is rate show the rate of each part against the total CPU time.

The rate of the number of vertices of the reduced subgraph against the original graph is called "reduction rate". The reduction rate in reduced subgraph column shows this value.

From the CPU time of the breakdown, it is found out that "proper reduction" part slightly needs time. In other words, most of time for generating a reduced subgraph is spent for computing st-shortest paths.

When $n$=16384, GR($k$-biDij) takes 8.8 seconds, and $k$-biDij takes 4.6 seconds, so $k$-biDij outperforms GR($k$-biDij) by 1.9 times. On the other hand the search of the reduced subgraph takes only 0.38 seconds. This means that $k$-biDij can be speeded up by 12.3 times by GR if st-shortest paths are precomputed. In that case, we think all-pairs shortest paths are usually precomputed for other pairs of a source vertex and a target one. That is, although GR($k$-biDij) is overperformed by $k$-biDij at first sight, $k$-biDij can be further speeded up by GR and precomputing all-pairs shortest paths.

When $n$=16384, the reduction rate is $n$=16384, and the original graph is reduced to 1/22.

Table 4.1 Comparison in CPU time in hypercube-shaped graphs

| n | CPU time(s) | | | | | |
|---|---|---|---|---|---|---|
| | GR(k-biDij) | k-biDij | GR(k-Dij) | k-Dij | k-biDij/ GR(k-biDij) | k-Dij/ GR(k-Dij) |
| 64 | 0.014 | 0.009 | 0.023 | 0.028 | 0.64 | 1.2 |
| 256 | 0.076 | 0.049 | 0.097 | 0.276 | 0.64 | 2.8 |
| 1024 | 0.354 | 0.246 | 0.519 | 3.814 | 0.69 | 7.3 |
| 4096 | 1.791 | 0.799 | 2.930 | 78.284 | 0.45 | 26.7 |
| 16384 | 8.807 | 4.630 | 14.703 | 5370.475 | 0.53 | 365.3 |

Table 4.2 CPU time breakdown of GR($k$-biDij) and reduced subgraph

| n | CPU time breakdown of GR(k-biDij) | | | | | | reduced subgraph | |
|---|---|---|---|---|---|---|---|---|
| | st-shortest paths | | proper reduction | | subgraph search | | numbe of vertices | reduction rate |
| | time(s) | rate | time(s) | rate | time(s) | rate | | |
| 64 | 0.009 | 0.391 | 0.000 | 0.000 | 0.004 | 0.174 | 26 | 0.406 |
| 256 | 0.052 | 0.536 | 0.003 | 0.031 | 0.022 | 0.227 | 76 | 0.297 |
| 1024 | 0.297 | 0.572 | 0.006 | 0.012 | 0.052 | 0.100 | 161 | 0.157 |
| 4096 | 1.618 | 0.552 | 0.017 | 0.006 | 0.156 | 0.053 | 379 | 0.093 |
| 16384 | 8.387 | 0.570 | 0.044 | 0.003 | 0.376 | 0.026 | 740 | 0.045 |

Table 4.3 CPU time breakdown of GR($k$-Dij) and reduced subgraph

| n | CPU time breakdown of GR(k-Dij) | | | | | | reduced subgraph | |
|---|---|---|---|---|---|---|---|---|
| | st-shortest paths | | proper reduction | | subgraph search | | numbe of vertices | reduction rate |
| | time(s) | rate | time(s) | rate | time(s) | rate | | |
| 64 | 0.017 | 0.739 | 0.000 | 0.000 | 0.005 | 0.217 | 26 | 0.406 |
| 256 | 0.051 | 0.526 | 0.003 | 0.031 | 0.044 | 0.454 | 76 | 0.297 |
| 1024 | 0.299 | 0.576 | 0.006 | 0.012 | 0.215 | 0.414 | 161 | 0.157 |
| 4096 | 1.613 | 0.551 | 0.017 | 0.006 | 1.300 | 0.444 | 379 | 0.093 |
| 16384 | 8.420 | 0.573 | 0.044 | 0.003 | 6.239 | 0.424 | 740 | 0.045 |

4.2 Comparison in scale-free graphs

The result of comparison in sparse case ($n' = 2$) is mentioned in 4.2.1, and that in dense case ($n' = \sqrt{n}$) is mentioned in 4.2.2.

4.2.1 Sparse case ($n' = 2$)

Fig. 4.2 and Table 4.4 shows the comparison in CPU time of four algorithms. When $n = 16384$, GR($k$-biDij) is outperformed by $k$-biDij by 1.36 times, and it is difficult to judge whether the speed rate increases or decreases as $n$ increases from only these data. GR($k$-Dij) outperforms $k$-Dij by 4.5 times, and the speed rate increases as $n$ is increases.

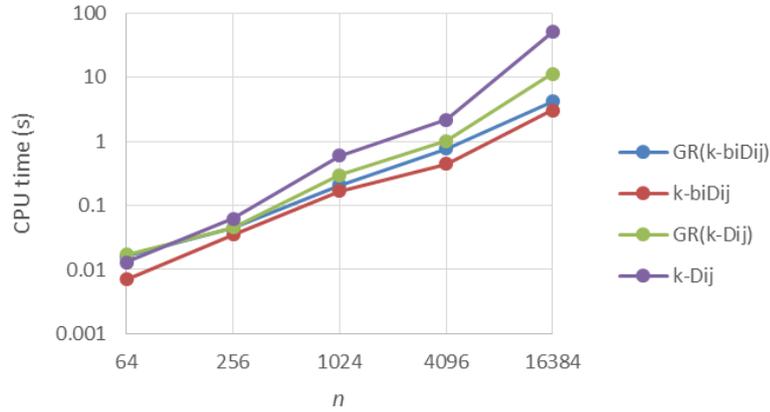

Fig. 4.2 Comparison in CPU time of four algorithms in sparse scale-free graphs

Table 4.4 Comparison in CPU time of four algorithms in sparse scale-free graphs

| n | CPU time(s) | | | | k-biDij/ GR(k-biDij) | k-Dij/ GR(k-Dij) |
|---|---|---|---|---|---|---|
|   | GR(k-biDij) | k-biDij | GR(k-Dij) | k-Dij | | |
| 64 | 0.016 | 0.007 | 0.017 | 0.013 | 0.44 | 0.8 |
| 256 | 0.044 | 0.035 | 0.045 | 0.062 | 0.80 | 1.4 |
| 1024 | 0.207 | 0.167 | 0.297 | 0.596 | 0.81 | 2.0 |
| 4096 | 0.768 | 0.445 | 1.007 | 2.145 | 0.58 | 2.1 |
| 16384 | 4.157 | 3.045 | 11.229 | 50.969 | 0.73 | 4.5 |

Table 4.5 CPU time breakdown of GR($k$-biDij) and reduced subgraph

| n | CPU time breakdown of GR(k-biDij) | | | | | | reduced subgraph | |
|---|---|---|---|---|---|---|---|---|
|   | st-shortest paths | | proper reduction | | subgraph search | | numbe of vertices | reduction rate |
|   | time(s) | rate | time(s) | rate | time(s) | rate | | |
| 64 | 0.010 | 0.588 | 0.001 | 0.059 | 0.005 | 0.294 | 43 | 0.672 |
| 256 | 0.025 | 0.556 | 0.001 | 0.022 | 0.017 | 0.378 | 64 | 0.250 |
| 1024 | 0.120 | 0.404 | 0.007 | 0.024 | 0.080 | 0.269 | 238 | 0.232 |
| 4096 | 0.545 | 0.541 | 0.023 | 0.023 | 0.200 | 0.199 | 637 | 0.156 |
| 16384 | 2.529 | 0.225 | 0.110 | 0.010 | 1.519 | 0.135 | 2717 | 0.166 |

Table 4.6 CPU time breakdown of GR($k$-Dij) and reduced subgraph

| n | CPU time breakdown of GR(k-Dij) | | | | | | reduced subgraph | |
|---|---|---|---|---|---|---|---|---|
|   | st-shortest paths | | proper reduction | | subgraph search | | numbe of vertices | reduction rate |
|   | time(s) | rate | time(s) | rate | time(s) | rate | | |
| 64 | 0.012 | 0.706 | 0.001 | 0.059 | 0.004 | 0.235 | 43 | 0.672 |
| 256 | 0.030 | 0.667 | 0.001 | 0.022 | 0.014 | 0.311 | 64 | 0.250 |
| 1024 | 0.118 | 0.397 | 0.007 | 0.024 | 0.171 | 0.576 | 238 | 0.232 |
| 4096 | 0.541 | 0.537 | 0.022 | 0.022 | 0.443 | 0.440 | 637 | 0.156 |
| 16384 | 2.534 | 0.226 | 0.109 | 0.010 | 8.586 | 0.765 | 2717 | 0.166 |

Tables 4.5 and 4.6 show the breakdown of CPU time of GR($k$-biDij) and GR($k$-Dij) and the number of vertices of the reduced subgraph and its rate respectively.

The proper reduction part slightly needs time like hypercube-shaped graphs.

When $n$=16384, GR($k$-biDij) takes 4.2 seconds, and $k$-biDij takes 3.0 seconds, so $k$-biDij outperforms GR($k$-biDij) by 1.4 times. On the other hand the search of the reduced subgraph takes 1.52 seconds. This means that $k$-biDij can be speeded up by 2.0 times by GR if the all-pairs shortest paths are precomputed. GR($k$-Dij) outperforms $k$-Dij by 4.5 times.

When $n$=16384, the reduction rate is 0.166, and the original graph is reduced to 1/6. In case of a hypercube-shaped graph, it is reduced to 1/22, so a graph is reduced more in hypercube-shaped graphs.

4.2.2 Dense case ($n' = \sqrt{n}$)

Fig. 4.3 and Table 4.7 shows the comparison in CPU time of four algorithms. When $n = 16384$, GR($k$-biDij) overperforms $k$-biDij by 2.6 times, and the speed rate increases as $n$ increases. GR($k$-Dij) outperforms $k$-Dij by 230 times, and the speed rate increases as $n$ is increases likewise.

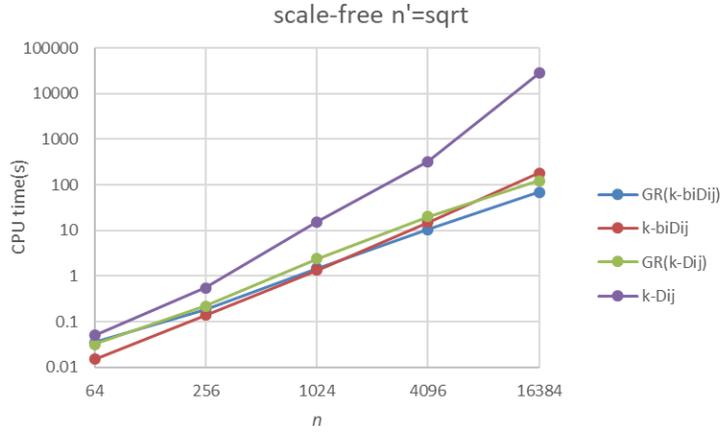

Fig. 4.3 Comparison in CPU time of four algorithms in dense scale-free graphs

Tables 4.8 and 4.9 show the breakdown of CPU time of GR($k$-biDij) and GR($k$-Dij) and the number of vertices of the reduced subgraph and its rate respectively.

The proper reduction part more slightly needs time than sparse scale-free graph, because the smaller the reduced subgraph is, the less time the proper reduction part of GR seems to tend to need.

When $n$=16384, GR($k$-biDij) takes 69 seconds, and $k$-biDij takes 183 seconds, so GR($k$-biDij) outperforms $k$-biDij by 2.7 times. On the other hand, the search of the reduced subgraph takes 5.2 seconds. This means that $k$-biDij can be speeded up by 35 times by GR and GR($k$-biDij) can be speeded up by 13 times if the all-pairs shortest paths are precomputed. GR($k$-Dij) outperforms $k$-Dij by 230 times.

When $n$=16384, the reduction rate is 0.056, and the original graph is reduced to 1/18. That is, the graph is reduced by 3.0 times in comparison with the sparse scale-free graph.

In case of sparse scale-free graphs, GR($k$-biDij) is outperformed by $k$-biDij by 1.4 times, and GR($k$-Dij) outperforms $k$-Dij by only 4.5 times. On the other hand, in case of dense scale-free graphs, GR($k$-biDij) outperforms $k$-biDij by 2.7 times and GR($k$-Dij) outperforms $k$-Dij by 230 times. The reason why GR performance is good in dense scale-free graphs seems to be that the reduction rate is 3.0 times smaller than sparse scale-free graphs.

Table 4.7 Comparison in CPU time of four algorithms in dense scale-free graphs

| n | CPU time(s) | | | | k-biDij/ GR(k-biDij) | k-Dij/ GR(k-Dij) |
|---|---|---|---|---|---|---|
| | GR(k-biDij) | k-biDij | GR(k-Dij) | k-Dij | | |
| 64 | 0.035 | 0.015 | 0.032 | 0.050 | 0.43 | 1.6 |
| 256 | 0.186 | 0.138 | 0.217 | 0.543 | 0.74 | 2.5 |
| 1024 | 1.472 | 1.327 | 2.394 | 15.539 | 0.90 | 6.5 |
| 4096 | 10.521 | 15.035 | 20.264 | 325.581 | 1.43 | 16.1 |
| 16384 | 69.159 | 182.523 | 122.799 | 28278.155 | 2.64 | 230.3 |

Table 4.8 CPU time breakdown of GR($k$-biDij) and reduced subgraph

| n | CPU time breakdown of GR(k-biDij) | | | | | | reduced subgraph | |
|---|---|---|---|---|---|---|---|---|
| | st-shortest paths | | proper reduction | | subgraph search | | numbe of vertices | reduction rate |
| | time(s) | rate | time(s) | rate | time(s) | rate | | |
| 64 | 0.026 | 0.813 | 0.000 | 0.000 | 0.008 | 0.250 | 26 | 0.406 |
| 256 | 0.152 | 0.700 | 0.002 | 0.009 | 0.032 | 0.147 | 51 | 0.199 |
| 1024 | 1.181 | 0.493 | 0.012 | 0.005 | 0.278 | 0.116 | 183 | 0.179 |
| 4096 | 8.640 | 0.426 | 0.063 | 0.003 | 1.817 | 0.090 | 540 | 0.132 |
| 16384 | 63.728 | 0.519 | 0.224 | 0.002 | 5.207 | 0.042 | 916 | 0.056 |

Table 4.9 CPU time breakdown of GR($k$-Dij) and reduced subgraph

| n | CPU time breakdown of GR(k-Dij) | | | | | | reduced subgraph | |
|---|---|---|---|---|---|---|---|---|
| | st-shortest paths | | proper reduction | | subgraph search | | numbe of vertices | reduction rate |
| | time(s) | rate | time(s) | rate | time(s) | rate | | |
| 64 | 0.019 | 0.594 | 0.000 | 0.000 | 0.012 | 0.375 | 26 | 0.406 |
| 256 | 0.157 | 0.724 | 0.002 | 0.009 | 0.057 | 0.263 | 51 | 0.199 |
| 1024 | 1.178 | 0.492 | 0.012 | 0.005 | 1.204 | 0.503 | 183 | 0.179 |
| 4096 | 8.608 | 0.425 | 0.064 | 0.003 | 11.591 | 0.572 | 540 | 0.132 |
| 16384 | 63.686 | 0.519 | 0.230 | 0.002 | 58.882 | 0.479 | 916 | 0.056 |

4.3 Consideration

We show our consideration on the result of comparison as follows:
1) Comparison of CPU time
- Comparison of GR($k$-biDij) with $k$-biDij

When $n = 16384$, GR($k$-biDij) outperforms $k$-biDij by 0.53, 0.79, and 2.64 times respectively in hypercube-shaped, sparse scale-free and dense scale-free graphs. That is $k$-biDij outperforms GR($k$-biDij) in hypercube-shaped, sparse scale-free graphs, and conversely GR($k$-biDij) outperforms $k$-biDij in dense scale-free graphs.

On the speed rate, it is difficult to judge whether it increases or decreases as $n$ increases from only the data in cases of hypercube-shaped and sparse scale-free graphs, but in case of dense scale-free graphs, it increases as $n$ increases, and it is forecast that it increases further as $n$ increases further.

- Comparison of GR($k$-Dij) with $k$-Dij

When $n = 16384$, GR($k$-Dij) outperforms $k$-Dij by 365, 4.5 and 230 times respectively in hypercube-shaped, sparse scale-free and dense sparse scale-free graphs. The speed rate increases as $n$ increases, and it is forecast that it increases further as $n$ increases further.

2) On the relationship between reduction rate and speed rate.

On the reduction rate of GR($k$-Dij) and GR($k$-biDij), it is 0.045, 0.166, and 0.056 respectively when $n = 16384$ in case of hypercube-shaped, sparse scale-free, and dense scale-free graphs. The reduction rate is independent of $E_x$ of GR($E_x$). On the other hand, the speed rate of GR($k$-Dij) is 365, 4.5, and 230 respectively when $n = 16384$ in the three graphs in the same order. Therefore, we can see that the smaller the reduction rate is, the larger the speed rate is.

In sparse scale-free graphs, the speed rate is 4.5 when $n = 16384$, and that is much worse than two other graphs. The reason seems to be that the reduction rate is 0.166 and more than three times larger than those of two other graphs.

3) On the effectiveness of precomputing all-pairs shortest paths

Precomputing all-pairs shortest paths makes it unnecessary to compute st-shortest paths in GR mentioned above. We summarize the effectiveness when $n = 16384$. $K$-biDij can be speeded up by 12.3, 2.0 and 35 times respectively in case of hypercube-shaped, sparse scale-free, and dense scale-free graphs. GR is effective even for $K$-biDij, which is the fastest as long as we know.

4) Graph reduction independent of $k$

Here we discuss the time complexity of graph reduction, that is, generating a reduced subgraph including computing st-shortest paths. As mentioned above, the time needed for proper graph reduction is very short. In other words, most of time for generating a reduced subgraph is spent for computing st-shortest paths. Therefore, if all-pairs shortest paths are precomputed and stored in main or secondary memory, graph reduction can be done quickly and the time complexity of graph reduction seems to be independent of $k$. The reason is as follows: The time complexity of computing st-shortest paths by Dijkstra algorithm is $O(m + n \log n)$ where $m$ is the number of edges. On the other hand, the time complexity of the part of proper graph reduction is $O(n \log n)$ because the heaviest part of it is sorting. In total, the time complexity of graph reduction is $O(m + n \log n)$, which is independent of $k$, not like $O(k(m + n \log n))$. This seems to be a good feature of GR.

5) Sparse graphs vs. dense graphs

From the result of the experiments using sparse and dense scale-free graphs, it is forecast that GR is more effective in dense graphs that sparse ones. However, it is effective in hypercube-shaped graphs which seem no to be so dense because the reduction rate is much smaller that sparse scale-free graphs. Therefore, we think it is necessary to analyze the behavior of GR further.

５．Conclusion

We proposed a new approach called GR algorithm for loop-less k-shortest path search. And we introduced the $k$-biDij which is the state-of-the-art algorithm and the fastest as long as we know.

The results mentioned above show the following:

1) GR($k$-biDij) outperforms $k$-biDij on CPU time in dense scale-free graphs, although $k$-biDij outperforms $k$-biDij in hypercube-shaped and sparse scale-free graphs.
2) By precomputing all-pairs shortest paths, GR always outperforms $k$-biDij on CPU time.
3) The time complexity of graph reduction is $(m + n \log n)$, which is independent of $k$.
4) The smaller the reduced subgraph is, the faster GR tends to run.
5) GR($k$-Dij), $k$-biDij, and GR($k$-biDij) outperform $k$-Dij on CPU time greatly.

References

[K-shortest]   Wikipedia's title: k shortest path routing.
[Yen71]        J. Y. Yen. Finding the k-Shortest Loopless Paths in a Network. Management Science. 17 (11): 712–716, 1971.
[Zhao14]       Z. Zhao and Y. Zong. An N-Shortest-Paths Algorithm Implemented with Bidirectional Search, An Interdisciplinary Response to Mine Water Challenges - Sui, Sun & Wang (eds), 2014.

Appendix: On our improvements to $k$-Dijkstra algorithm and $k$-bidirectional Dijkstra algorithm

We mainly added the following improvements: 1) and 2) are added to $k$-Dijkstra algorithm, and 1), 2), and 3) are added to $k$-bidirectional Dijkstra algorithm.

1) Pruning by counting the shortest paths at each vertex

As mentioned above, $k$-Dijkstra algorithm in [K-shortest] searches beyond a vertex $v$ if the number of paths from the source vertex is less than or equal to $k$. We improved further so that when searching beyond $v$, a new path $p_w$ which is a path from a starting vertex to $w$ adjacent to $v$ is added to a priority queue if the number of paths to $w$ from the source vertex is less than or equal to $k$.

2) Pruning using priority queues at each vertex

A priority que $Q$ is placed at a starting vertex $s$ (a source vertex and a target vertex in case of $k$-bidirectional Dijkstra algorithm). The path from $s$ is enqueued into $Q$. It is desirable that the number of paths enqueued to it is as small as possible because the cost to maintain the order becomes smaller. When a path $p_v$ which is a path from $s$ $v$ is dequeued from $Q$, a vertex $w$ adjacent to $v$ is traversed. Then $p_w$ is enqueued to $Q$ if and only if $p_w$ is shorter than or equal to the $k$-th shortest path at $w$ using the priority queue at each vertex. The priority queue only keeps the top-$k$ shortest paths. This improvement reduces the number of vertices enqueued to $Q$.

3) Pruning by a terminating condition

In our implementation, when at a vertex $v$, a path from $p_v^s$ from $v_s$ to $v$, and $p_v^t$ from $v_t$ to $v$ conflicts, a path $p$ is composed from $p_v^s$ and $p_v^t$, and it is enqueued into a priority queue $Q$. $Q$ only keeps the top-$k$ shortest paths. When $Q$ is full, we utilize the value $l_k$ which is the length of the $k$-th shortest path in $Q$. Consider a case that a vertex $v$ is visited. If the shortest path $p_v^t$ whose length is $l$ from $v_t$ to $v$ is already generated, it is meaningless to traverse beyond $v$ if $l_k \leq len(p_v)+ l$ because if at $x$ beyond $v$ a path whose length is less than $l_k$ is composed, it means there is a shorter path from $v_t$ to $v$ than the shortest path $p_v^t$. Therefore, traversing beyond $v$ is stopped, that is, pruning is done. Otherwise let the path at the top of $Q_t$ be expressed by $p_1^t$. If $l_k \leq len(p_v)+ len(p_1^t)$, there is not a possibility that a path less than or equal to $l_k$ is generated at $v$, so pruning is done.

We are planning to submit a paper which describes the details in the near future.